\documentclass[aps,prb,twocolumn,superscriptaddress]{revtex4-1}

\usepackage{amsmath}
\usepackage{rotating}
\usepackage{graphicx}
\usepackage{hyperref}
\usepackage{epstopdf}
\usepackage{color}
\usepackage{booktabs}
\usepackage{longtable}
\usepackage{float}

\usepackage{color}
\usepackage{colortbl}
\usepackage{natbib}
\usepackage{hyperref}

\definecolor{PB}{rgb}{0,0,0}
\definecolor{changes}{rgb}{0,0,0}
\definecolor{changes2}{rgb}{0.0,0.0,0.0}
\definecolor{changes3}{rgb}{0,0,0}
\definecolor{changes5}{rgb}{0,0,0}
\definecolor{changes4}{rgb}{0,0,0}
\definecolor{changes6}{rgb}{0,0,0}
\definecolor{changes7}{rgb}{0,0,0}
\definecolor{lightgray}{rgb}{0.72,0.72,0.72}

\begin{document}


\title{The interplay between strong correlation and adsorption distances}


\author{Marc Philipp Bahlke}
\affiliation{Institut f\"ur Anorganische und Angewandte Chemie, Universit\"at Hamburg, Martin-Luther-King-Platz 6, 20146 Hamburg, Germany}
\author{Michael Karolak}
\affiliation{Institut f\"ur Theoretische Physik und Astrophysik, Universit\"at W\"urzburg, Am Hubland, 97074 W\"urzburg, Germany}
\author{Carmen Herrmann}
\affiliation{Institut f\"ur Anorganische und Angewandte Chemie, Universit\"at Hamburg, Martin-Luther-King-Platz 6, 20146 Hamburg, Germany}


\date{\today}

\begin{abstract}
\textcolor{changes4}{Adsorbed transition metal atoms can have partially filled $d$- or $f$-shells due to strong on-site Coulomb interaction. Capturing all effects originating from electron correlation in such strongly correlated systems is a challenge for electronic structure methods. \textcolor{changes5}{It requires a sufficiently accurate description of the atomistic structure (in particular bond distances and angles), which is usually obtained from \textcolor{PB}{first-principles} Kohn--Sham density functional theory (\textcolor{PB}{DFT}), which due to the approximate nature of the exchange-correlation functional may provide an unreliable description of strongly correlated systems. To elucidate the consequences of this popular procedure}, we apply a combination of DFT with the Anderson impurity model (AIM), \textcolor{changes5}{as well as} DFT+$U$ for a calculation of the potential energy surface along the Co/Cu(001) adsorption coordinate, and compare the results with those obtained from DFT. The adsorption minimum is shifted 
towards larger distances by applying DFT+AIM, or the much cheaper DFT$+U$ method, compared to the corresponding spin-polarized 
DFT results, by a magnitude comparable to variations between different approximate exchange-correlation functionals (0.08 to 0.12 {\AA}). This shift originates from an increasing correlation energy at larger adsorption distances, \textcolor{changes5}{which can be traced back to the Co $3d_{xy}$ and  $3d_{z^2}$ orbitals being more correlated as the adsorption distance is increased.}} We can show that such considerations are important, as they may strongly affect electronic properties such as the Kondo temperature.
\end{abstract}


\maketitle



\section{Introduction}\label{intro}
Strongly correlated \textcolor{changes4}{systems} have gained much interest in recent research due to a variety of interesting phenomena, such as high-temperature superconductivity, colossal magneto-resistance, or the zero-bias anomaly in scanning-tunneling microscopy experiments \cite{Dagotto2005}. These phenomena originate from the interplay of localized $d$- or $f$-electrons (magnetic impurities) \textcolor{changes6}{with their} surrounding electrons, and are considerably sensitive to temperature, pressure and doping \cite{Kotliar2004}. 
Often first-principles investigations are required to understand such effects, which have shown to be challenging \cite{Lueders2005, Marques2005, Lichtenstein1998} with the standard  Kohn--Sham density functional theory formalism (referred to  as DFT in the following) \cite{hohenbergkohn, Kohn1965}.
Although being formally exact, \textcolor{PB}{Kohn--Sham} DFT suffers in practice  from the lack of knowledge on the exact exchange--correlation functional, and its application to strongly correlated materials can thus give inaccurate results, as in the case of, e.g., Mott insulators \cite{Anisimov1991}. On the other hand, DFT has been successfully applied in quantum chemistry for describing the electronic ground states and \textcolor{changes}{molecular structures} of transition-metal complexes, \textcolor{changes7}{systems which can have both dynamical and static correlation} (\textcolor{changes2}{for a short explanation of both terms, see Sec.\ \ref{dyn_stat}}), with the results strongly depending on the choice of the exchange--correlation functional \cite{Cramer2006, Buehl2008, Cramer2009}.\\
\textcolor{changes2}{The goal of first-principles calculations is to (1) describe the potential energy surface (PES), i.e. structural parameters, in a correct manner, and (2)} \textcolor{changes}{ to explain \textcolor{PB}{and predict} experimentally observed electronic properties, such as the ones resulting from strong correlation  mentioned before.} Both tasks are usually \textcolor{changes4}{combined} by first optimizing a \textcolor{changes2}{molecular or adsorbate structure \textcolor{PB}{within Kohn--Sham DFT employing} an appropriate exchange--correlation functional, and then using an electronic structure method that can describe correlation more accurately. In solid-state physics, this is often a method which \textcolor{PB}{relies on} certain input from DFT.} One approach which has been satisfactorily applied to overcome, \textcolor{changes2}{e.g.}, the band gap problem \cite{Seidl1996} of strongly correlated materials in DFT is known as DFT$+U$ \cite{Anisimov1991, Liechtenstein1995}, in which a simple on-site Coulomb potential ($U$) is added to the orbitals of interest (e.g., the $d$-orbitals). The advantage of this method is the simplicity and the low computational cost. DFT$+U$ suffers from the neglect of dynamical effects \textcolor{changes7}{in the sense used in solid-state physics (see Sec.\ \ref{dyn_stat}), and is thus not applicable to all strongly correlated systems for the explanation of experimental observations as, e.g., those which exhibit the Kondo effect.}
\textcolor{changes3}{In this direction, the combination of DFT with the Anderson impurity model (AIM) \cite{ande61}, whose solution explicitly contains the electron correlation effects on a magnetic impurity (capturing also dynamical electron correlation effects \textcolor{changes4}{in the physics sense}),  has been employed very successfully, both in the context of impurities and bulk materials within dynamical mean field theory (DMFT)\cite{dmftrev96}.} This ultimately allows for the extraction of dynamical properties as, \textcolor{changes7}{e.g., the frequency-dependent self-energy}, which in turn are necessary to explain experimental observations in more detail. \textcolor{changes2}{In addition, the electron density of the system resulting from the solution of the AIM  may differ somewhat from DFT and may thus change its energy.\\
One may  ask how strongly  the solution of the AIM (more precisely its combination with DFT, further called DFT++, \textcolor{changes4}{following Lichtenstein \textit{et al.}}\cite{Lichtenstein1998} - see next section) affects potential energy surfaces compared to choosing different approximate exchange correlation functionals: If the effect of DFT vs. DFT++ is on the same order of magnitude or smaller than the effect of varying the exchange--correlation functional, it may be justified to continue the common practice of optimizing structures with DFT \textcolor{changes7}{provided a suitable exchange--correlation functional is carefully chosen}. From previous work, it is known that that the difference of the predicted lattice parameters between LDA+$U$ and LDA in combination with dynamical mean field theory\footnote{In general, DFT+DMFT belongs to the so-called class of DFT++ approaches, but here we strictly distinguish between DFT++ and DFT+DMFT, because in the DMFT cycle the model Hamiltonian (as, e.g., the Anderson 
Hamiltonian) is solved iteratively, which is required for treating systems with periodically repeating impurities.} (LDA+DMFT, using the Hubbard I solver)  of Ce$_2$O$_3$ and Pu$_2$O$_3$ do agree with each other, if the DMFT cycle is done over charge self-consistency \cite{amadon13}.} Furthermore, the results of Skornyakov \textit{et al.}\cite{Skornyakov2017} show that the lattice parameter of FeSe shifts by roughly +0.1 Bohr (0.05~\AA) by using PBE in combination with DMFT \textcolor{changes7}{using the continuous-time quantum Monte Carlo method  as impurity solver,} compared to a non-magnetic PBE calculation. \textcolor{changes4}{It is unclear whether there are similarily small effects for strongly correlated adsorbates on surfaces.}\\
\textcolor{changes}{To this end, we apply DFT++  to compute the PES of Co/Cu(001) for a scan along the Co-surface distance, and in addition, DFT$+U$ for optimizing the Co adsorption distance on Cu(001).} This system has been under extensive investigation in tunneling microscopy experiments, in which a zero-bias anomaly was detected due to the Kondo effect (with a Kondo temperature $T_K$ of 88 $\pm$ 4 K) \cite{wahl02,wahl04,wahl09, Neel2007}. 
Since the Kondo effect results from the correlation between conduction band electrons and localized unpaired electrons of a single atom or molecule, the application of the AIM in theoretical investigations should account for this effect, as it was shown earlier by Jacob (one crossing approximation) \cite{jaco15} and Baruselli \textit{et al}. (numerical renormalization group) \cite{baru15}. \textcolor{changes3}{However, we focus here on the PES along the Co on Cu(001) adsorption coordinate, and give only a qualitative estimate of the Kondo temperature.} Further motivation for the choice of Co/Cu(001) does not rely only on the simplicity of the system, but also on the exciting opportunity to control the Kondo effect via the formation of carbonyl-cobalt complexes by attaching CO ligands. 
For instance, it was shown that the  Kondo temperature of a Co(CO)$_4$ molecule on Cu(001) is about $T_K$ = 283 $\pm$ 36 K \cite{Wahl2005}. \\
In this work, we show that the energetic minimum of the  Co/Cu(001) adsorption distance \textcolor{changes4}{is shifted by 0.08~\AA~- 0.12~\AA~towards larger values if one applies DFT++, or the much cheaper DFT$+U$ method, compared to the corresponding spin-polarized} DFT calculations. Our results suggest that \textcolor{changes4}{even though larger than what has been reported for solids,} the effect of dynamical correlation \textcolor{changes2}{(i.e., the energy-dependent self-energy taken into account by DFT++, but not by DFT$+U$)} \textcolor{changes5}{leads to a correction of the adsorption distance  of the same order as changing the approximate exchange--correlation functional of \textcolor{PB}{Kohn--Sham} DFT}. Nonetheless, we point out here that dynamical \textcolor{changes2}{properties of DFT++} are important to describe the electronic structure of many strongly correlated materials. Further, \textcolor{PB}{electronic properties related to correlation may depend on the surface--adsorbate distance, as has been pointed out for model systems in the context of energy transfer and electronic friction~\cite{Plihal1998}. We} use the  results of the simplest Kondo model \textcolor{changes7}{which is the one-band case with a flat bath \cite{hewson}} to approximate the Kondo temperatures 
based on our DFT parameters at different points of the PES, \textcolor{changes}{to study the sensitivity of approximated Kondo temperatures to adsorption distances based on DFT calculations.}
\section{Methodology}\label{method}
\begin{figure} [h]
\centering
\includegraphics[width=0.5\textwidth]{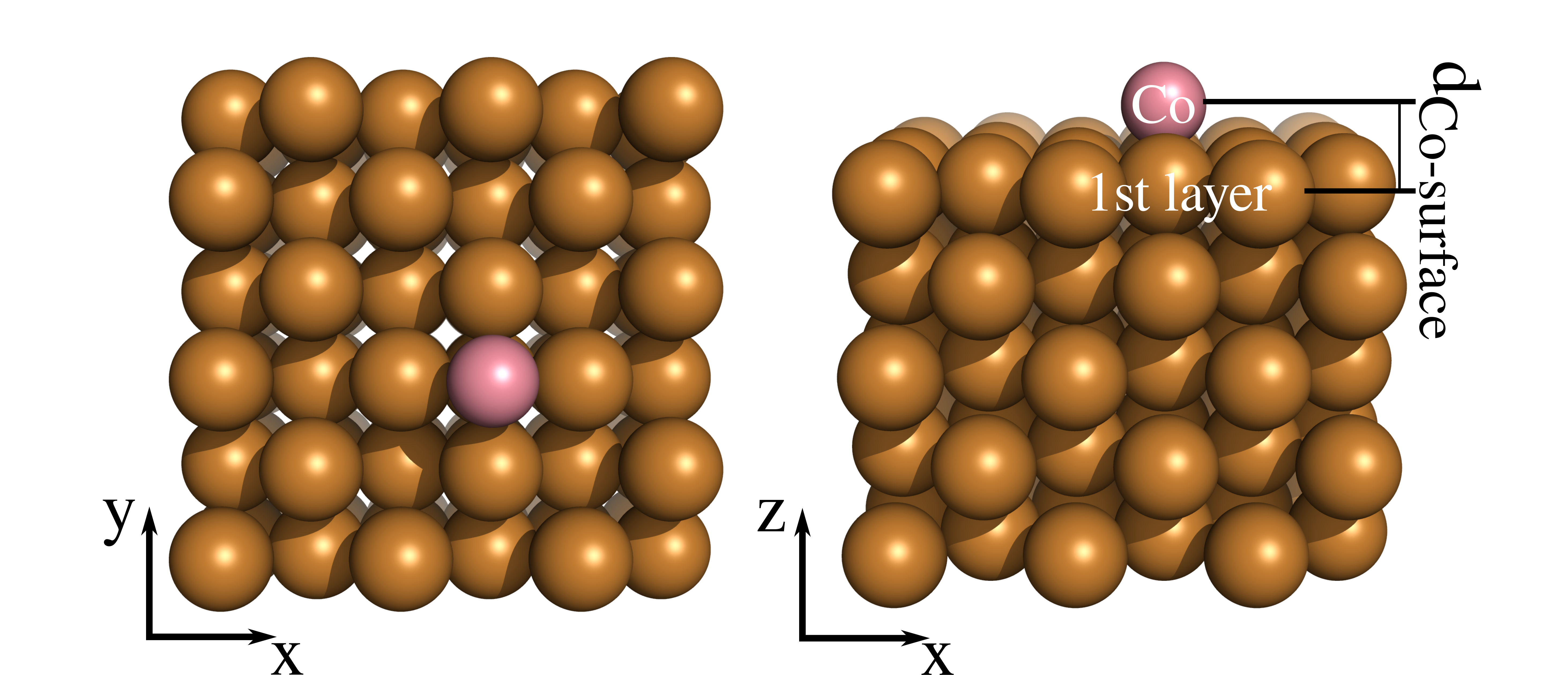}
\caption{Unit cell used for calculating the PES for Co/Cu(001) adsorption. \textcolor{changes2}{$d_{\mathrm{Co-surface}}$ is the difference between the $z$-coordinate of the Cu atoms in the first layer (they were aligned so that all Cu atoms in the first layer have a $z$-coordinate of 0.00) and the $z$-coordinate of the Co atom.}}
\label{unitcell}
\end{figure}

We modeled the Cu(001) surface by a 4 x 4 super cell \textcolor{changes7}{with a lattice parameter of 3.615~\AA~\cite{Wyckoff}} and placed the Co atom in the fourfold-hollow position (see Fig.\ \ref{unitcell}), resulting in an interatomic distance between neighboring Co atoms of 10.8~\AA. The PES was computed by doing single-point \textcolor{changes4}{electronic structure} calculations at Co--surface distances ($d_{\mathrm{Co-surface}}$, see Fig.\ \ref{unitcell}) between 1.3~\AA~and 1.7~\AA~\textcolor{changes4}{in increments of 0.10~\AA, with a finer spacing of 0.02~\AA~around the energetic minimum}. To avoid interactions between different unit cells in $z$-direction, we applied a vacuum of 10.0~\AA~on each unit cell (distance between the Co atom of a unit cell to the lowest Cu layer of the next unit cell in $z$-direction).\\
The single-point calculations (DFT, DFT$+U$ and DFT++) were done with the  \textsc{Abinit} 7.10.4 program package \cite{abinit} using a projector augmented plane wave (PAW) basis set \cite{blochl94, abinitpaw}. We applied the local density approximation (LDA) for the exchange--correlation functional of Perdew and Zunger \cite{pz81,ca80} and the generalized gradient approximation (GGA) of Perdew \textit{et al.} \cite{perd96}, both with and without spin polarization. 
For the plane-wave basis set, a cutoff of 544.2~eV for the kinetic energy and 2180~eV for the PAW double grid were chosen. As the convergence parameter for the self-consistent field algorithm, the tolerance on the total energy difference was set to $2.7\cdot10^{-7}$~eV. For the smearing of the metallic occupations, the Fermi-Dirac method was used with a smearing value of 0.01~eV.
For the $k$-points, a Monkhorst-Pack grid \cite{monkhorst} of size 4$\times$4$\times$1 was used. \\
For answering the question of how the PES for Co adsorption on a Cu(001) surface  is affected by taking into account strong electron correlation, we explicitly accounted for it on a local subspace spanned by the Co $3d$ orbitals. In the \textsc{Abinit} program package \cite{abinit}, the local basis is obtained by  projecting the Bloch states $\Psi_{\nu}$ onto a subset of localized orbitals $\chi_{i}$ centered on the correlated atom (for details about the implementation, see Ref. \onlinecite{Amadon2008}). This allows for the extraction of proper Wannier functions which can be used to describe the Co adatom within the AIM. In this work, we used the wave function obtained from spin-unpolarized DFT for the projection scheme including all occupied and 100 empty bands.\\
The Anderson Hamiltonian in its general multi-orbital form is written as
\begin{widetext}
\begin{equation}\label{siam_hamilton}
\hat{H}  = \sum_{\nu\sigma}\epsilon_{\nu}\hat{c}^{\dag}_{\nu,\sigma}\hat{c}_{\nu,\sigma} + \sum_{\nu i\sigma} [V_{\nu i}\hat{c}^{\dag}_{\nu,\sigma}\hat{d}_{i,\sigma} + V^*_{\nu i}\hat{d}^{\dag}_{i,\sigma}\hat{c}_{\nu,\sigma}] + 
\sum_{i\sigma}\epsilon_{i}\hat{d}^{\dag}_{i\sigma}\hat{d}_{i\sigma} + \frac{1}{2} \sum_{\substack{ijkl\\\sigma\sigma'}} U_{ijkl} \hat{d}^{\dag}_{i\sigma}\hat{d}^{\dag}_{j\sigma'}\hat{d}_{l\sigma'}\hat{d}_{k\sigma}.
\end{equation}
\end{widetext}
In Eq.\ (\ref{siam_hamilton}), $\epsilon_i$ is the energy of the $i$th localized $d$ orbital of the impurity (\textcolor{changes}{in this work, the Co $3d$ shell}), and $\epsilon_\nu$ is the kinetic energy of the bath electron $\nu$ (\textcolor{changes}{in this work, the copper surface}). $\hat{c}_{\nu\sigma}$/$\hat{c}^{\dag}_{\nu\sigma}$ are creation and annihilation operators for electrons with spin $\sigma$ acting on the $\nu$'th bath state, whereas $\hat{d}_{i\sigma}$/$\hat{d}_{i\sigma}^{\dag}$ are the corresponding operators acting on the local orbital $i$. The bath electrons are coupled to the impurity via the hybridization $V_{\nu i}$, and $U_{ijkl} = \int drdr'\psi_i^*(r)\psi_j^*(r')\frac{e^2}{|r-r'|}\psi_k(r)\psi_l(r')$ is the local Coulomb interaction (we dropped the spin indices here for simplicity) as introduced by Slater \cite{slater1960}, \textcolor{changes4}{with $\psi_x$ ($x = i,j,k,l$) \textcolor{changes5}{being in general any} atom-centered basis function}. \textcolor{changes5}{We used here 
the parameters ($F^0$, $F^2$ and $F^4$, see below) as derived by Slater for hydrogen-type atomic orbitals.}\\
In our treatment the hybridization  is given as an energy-dependent function $\Delta_{ij}(\omega)$:
\begin{equation}\label{hyb_function}
\Delta_{ij}(\omega) = \sum_{\nu} \frac{V_{\nu i} V_{\nu j}^{*}}{\omega + i0^+ - \epsilon_{\nu}}.
\end{equation}
In practice we obtain the hybridization function from the DFT solution of the total system, via the local Green's function $g_{ij}(\omega)$,  
\begin{equation}\label{nonint_imp}
g_{ij}(\omega) = [(\omega + i0^+)\delta_{ij} - \epsilon_{ij} - \Delta_{ij}(\omega)]^{-1},
\end{equation}
with $\epsilon_{ij}$ \textcolor{changes4}{being the matrix elements of the Kohn--Sham operator in the local impurity basis (i.e. the Co $3d$ orbitals, \textcolor{changes5}{as obtained after projection onto localized atomic orbitals $\psi_x$, see below). In an orthogonal basis ($\delta_{ij} = 0$) the diagonal elements $\epsilon_{ii}$ can be considered as representing the splitting of the Co $3d$ orbitals due to the surrounding atoms, which is known from crystal-field theory, or equivalently from lattice models (e.g. the Hubbard model). The effect of all other atoms in the system are now fully captured by the energy dependent hybridization function, where the imaginary part ($\mathrm{Im}\Delta_{ij}(\omega)$) broadens $\epsilon_{ij}$, and the real part ($\mathrm{Re}\Delta_{ij}(\omega)$) acts as an energy-dependent shift of $\epsilon_{ij}$.} \\ 
The local Green's function can be obtained by projecting the Kohn--Sham Green's function onto the local subspace of the impurity as}
\begin{equation}\label{nonint_imp_proj}
g_{ij}(\omega) = \sum_{\nu,\nu'} \hat{P}^{\phantom{}}_{i\nu}G^{KS}_{\nu\nu'}(\omega)\hat{P}^{\phantom{}}_{\nu'j},
\end{equation}
where
\begin{equation}\label{nonint_KS}
G^{KS}_{\nu\nu'} (\omega) = [(\omega + i0^+ + \mu - \epsilon_{\nu})\delta_{\nu\nu'}]^{-1}
\end{equation}
and
\begin{equation}\label{PROJ}
\hat{P}_{i\nu} = \langle\psi_{i}|\Psi_{\nu}\rangle,
\end{equation}
\textcolor{changes4}{with $\Psi_{\nu}$ being the the Kohn--Sham Bloch wavefunctions, and $\epsilon_{\nu}$  the corresponding Kohn--Sham eigenenvalues.}\\
\textcolor{changes}{Solving Eq.\ (\ref{nonint_imp}) for $\Delta_{ij}(\omega)$ yields}
\begin{equation}\label{hyb_function2}
\Delta_{ij}(\omega) = -[g_{ij}^{-1}(\omega) + \epsilon_{ij} - (\omega + 0^+)\delta_{ij}].
\end{equation}
\textcolor{changes}{We will use the energy-dependent hybridization function to solve the impurity model  with the continuous-time quantum Monte Carlo (CT-QMC) method in the hybridization expansion. The Coulomb part of Eq.\ (\ref{siam_hamilton}) \textcolor{changes7}{and for our DFT$+U$ calculations} is approximated by using only density-density terms ($i=k, l=j$ , yielding $\hat{n}_{i\sigma} = \hat{d}^{\dag}_{i\sigma^i}\hat{d}^{\phantomsection{}}_{i\sigma^i}$ and $\hat{n}_{j\sigma} = \hat{d}^{\dag}_{j\sigma^j}\hat{d}^{\phantomsection{}}_{j\sigma^j}$) of the full Coulomb matrix, \textcolor{changes4}{and setting all other matrix elements $U_{ijkl}$ to zero.}} Thus, the Coulomb part reduces to
\begin{equation}\label{Coulomb_dens-dens}
\begin{aligned}
\frac{1}{2}\sum_{\substack{ijkl\\\sigma\sigma'}}U_{ijkl}\hat{d}^{\dag}_{i\sigma}\hat{d}^{\dag}_{j\sigma'}\hat{d}_{l\sigma'}\hat{d}_{k\sigma} & \approx   \frac{1}{2}\sum_{\substack{ij\\\sigma\sigma'}} U_{ij} \hat{n}_{i\sigma}\hat{n}_{j\sigma'} \\
& + \frac{1}{2} \sum_{\substack{i\neq j\\\sigma}}  (U_{ij} - J_{ij})\hat{n}_{i\sigma}\hat{n}_{j\sigma}.
\end{aligned}
\end{equation}
$U_{ij}$ (\textcolor{changes4}{Coulomb integral}) and $J_{ij}$ (\textcolor{changes4}{exchange integral}) are the density-density terms of the full Coulomb matrix, \textcolor{changes4}{which read},

\begin{equation}\label{Coulomb}
\begin{aligned}
U_{ij} = \int drdr'\psi_i^*(r)\psi_j^*(r')\frac{e^2}{|r-r'|}\psi_i(r)\psi_j(r'),
\end{aligned}
\end{equation}
and
\begin{equation}\label{Exchange}
\begin{aligned}
J_{ij} = \int drdr'\psi_i^*(r)\psi_j^*(r')\frac{e^2}{|r-r'|}\psi_j(r)\psi_i(r').
\end{aligned}
\end{equation}
 We have obtained the full Coulomb interaction by the Slater integrals \cite{slater1960, Slater1929} $F^0$, $F^2$ and $F^4$ (for $3d$ systems), by choosing five different values for the average screened Coulomb interaction $U = F^0$ (3.0 eV, 3.8 eV, 4.0 eV, 4.2 eV and 5.0 eV) and $J = \frac{1}{14}(F^2+F^4) = 0.9$ eV with $\frac{F^4}{F^2} = 0.625$. \textcolor{changes}{The values $U$ = 4 eV and $J$ = 0.9 eV were taken from Ref. \onlinecite{surer11}, and to investigate the effect of $U$, we have varied its value between 3.0 eV and 5.0 eV.}\\
When using such combined approaches, one should be aware of the so-called ``double-counting'' of electron correlation, since some of the Coulomb interaction introduced with Eq.\ (\ref{siam_hamilton}) is already contained in the underlying KS-DFT. \textcolor{changes}{Two commonly used approaches for} calculating the double-counting energy are known as ``around mean field'' (AMF)\cite{Anisimov1991} and the ``fully localized limit'' (FLL)\cite{sawatzky94}. In this work we focus on the FLL method, which estimates the double-counting as 
\begin{equation}\label{fll_dc}
\begin{aligned}
E^{\mathrm{FLL}}_{\mathrm{DC}} = \frac{1}{2}UN(N-1) & - \frac{1}{2}JN_{\uparrow}(N_{\uparrow}-1) \\
& - \frac{1}{2}JN_{\downarrow}(N_{\downarrow}-1),
\end{aligned}
\end{equation}
 $N$ is the total number of electrons in the correlated subspace (\textcolor{changes}{the Co $3d$ shell in our case}), and $N_{\uparrow}$ and $N_{\downarrow}$ are the number of spin-up and spin-down electrons in the correlated subspace. \textcolor{changes}{$U$ ($F^0$) and $J$ ($\frac{1}{14}(F^2+F^4)$)} are the average Coulomb and exchange energies between the electrons in the correlated subspace. \\
\textcolor{changes}{After solving the AIM (Eq.\ (\ref{siam_hamilton})), one obtains the so-called interacting impurity Green's function $G^{\mathrm{IMP}}$ which is related to the non-interacting impurity Green's function ($g$) via the Dyson equation  }
\begin{equation}\label{dyson}
g^{-1}_{ij}(\omega) = G_{ij}^{\mathrm{IMP}^{-1}}(\omega) + \Sigma_{ij}(\omega),
\end{equation}
\textcolor{changes}{where all electron correlation effects within the AIM are contained in the so-called self-energy $\Sigma$ (for details about the concept of the self-energy, see e.g. the textbook of Coleman \cite{Coleman2015}).}\\
\textcolor{changes3}{For the calculation of the DFT++ energies, we used the dynamical mean field theory (DMFT) implementation of \textsc{Abinit}, doing only one single iteration of the DMFT cycle (\textcolor{changes4}{solving the AIM only once, because we have only one impurity (Co) coupled to a Cu(001) surface)}. The total energy within the DMFT framework is then given as \cite{amadon13}}
\begin{equation}\label{Edmft}
\begin{aligned}
 E_{\mathrm{DMFT}} &= T_0^{\mathrm{DFT+DMFT}} + E_{\mathrm{XC}} [\rho(\textbf{r})] - E_{\mathrm{Ha}} [\rho(\textbf{r})] \\
 &+ \int d\textbf{r} V_{\mathrm{ext}}(\textbf{r}) \rho(\textbf{r}) + E_{\mathrm{pot}} - E_{\mathrm{DC}}.
\end{aligned}
\end{equation}
In Eq.\ (\ref{Edmft}), $V_{\mathrm{ext}}$ is the external potential, $E_{\mathrm{XC}}$ is the exchange--correlation functional, and $E_{\mathrm{Ha}}$ is the Hartree energy (the Coulomb interaction between the electrons), $E_{\mathrm{pot}}$ is the on-site interaction term (Coulomb interaction on the impurity) and is directly calculated from the impurity solver (i.e. CT-QMC),  $E_{\mathrm{\mathrm{DC}}}$ is the double-counting correction, and  $T_0^{\mathrm{DFT+DMFT}}$ is the kinetic energy term that is an analogue to the DFT kinetic energy, but with a correction due to DMFT \textcolor{changes4}{(i.e. the occupation of the Kohn--Sham orbitals changes, and consequently the kinetic energy has to be corrected according to the new density obtained from DMFT)}. For further details about the implementation and total energy calculation, we refer to Ref. \onlinecite{amadon13}. As mentioned before, we are actually not using DMFT since we solve the AIM only once (calling it DFT++). Consequently in this 
work $E_{\mathrm{DFT++}}$ = $E_{\mathrm{DMFT}}$.\\
\textcolor{changes4}{The on-site potential energy $E_{\mathrm{pot}}$, as obtained from CT-QMC with density-density type interaction, is calculated as\cite{Gull2011}}
\begin{equation}\label{EPOT}
\begin{aligned}
E_{\mathrm{pot}} & = \frac{1}{2}\sum_{\substack{ij\\\sigma\sigma'}} U_{ij} \langle\hat{n}_{i\sigma}\hat{n}_{j\sigma'}\rangle  \\
&+ \frac{1}{2} \sum_{\substack{i\neq j\\\sigma}}  (U_{ij} - J_{ij}) \langle\hat{n}_{i\sigma}\hat{n}_{j\sigma}\rangle.
\end{aligned}
\end{equation}
\textcolor{changes4}{Here, $\langle \hat{n}_{i\sigma}\hat{n}_{j\sigma'}\rangle$ is the so-called double-occupancy.}
\section{Results and Discussion}\label{results}
 \textcolor{changes}{In the following  we are going to discuss the effect of strong electron correlation on the potential energy surface along the Co/Cu(001) adsorption coordinate $d_{\mathrm{Co-surface}}$ comparing DFT with  DFT++ and DFT+$U$ results.} The former are split in spin-unpolarized and spin-polarized calculations. We consider the spin-polarized calculations as our DFT reference, \textcolor{changes3}{ and for a \textcolor{changes4}{physically correct description} of the non-magnetic behavior of the copper surface, we use the spin-unpolarized solution as a starting point for the DFT++ calculations \textcolor{changes7}{where the magnetic moment of the Co $3d$ shell is described correctly by the AIM}, as is common in the literature\cite{jaco15}.}\\
After we have discussed the potential energy surfaces, we will consider the  electronic structure in more detail. We provide extensive data at selected values for $d_{\mathrm{Co-surface}}$ to show the sensitivity of the electronic properties as described by the AIM to the adsorption distance.  This allows us to judge the importance of accurate adsorption distances  for strongly correlated systems. As mentioned in the introduction, it was found that Co on Cu(001) exhibits a Kondo effect with a Kondo temperature of $88 \pm 4$ K \cite{wahl02,wahl04,wahl09}. In this work, we will not focus on this temperature in detail, because  the largest inverse temperature $\beta = 100$~eV$^{-1}$ employed here  corresponds to $\sim 116$~K which is too high to observe  a Kondo effect in our calculations.  \footnote{At lower temperatures (e.g. $\beta = 200$ eV$^{-1}$ ($\sim 58$~K)) one would need a sufficiently large $k$-point grid for a proper integration of the Greens function resulting in a computational effort 
that goes beyond the scope of this work}. We will use our DFT results for a qualitative assessment of the Kondo physics, as will be discussed in part B of this section. \textcolor{changes7}{A quantitative analysis will be will be the subject of future work.}
\subsection{Potential energy surface scan along the adsorption coordinate}\label{pes}
\begin{figure} [H]
\centering
\includegraphics[width=0.5\textwidth]{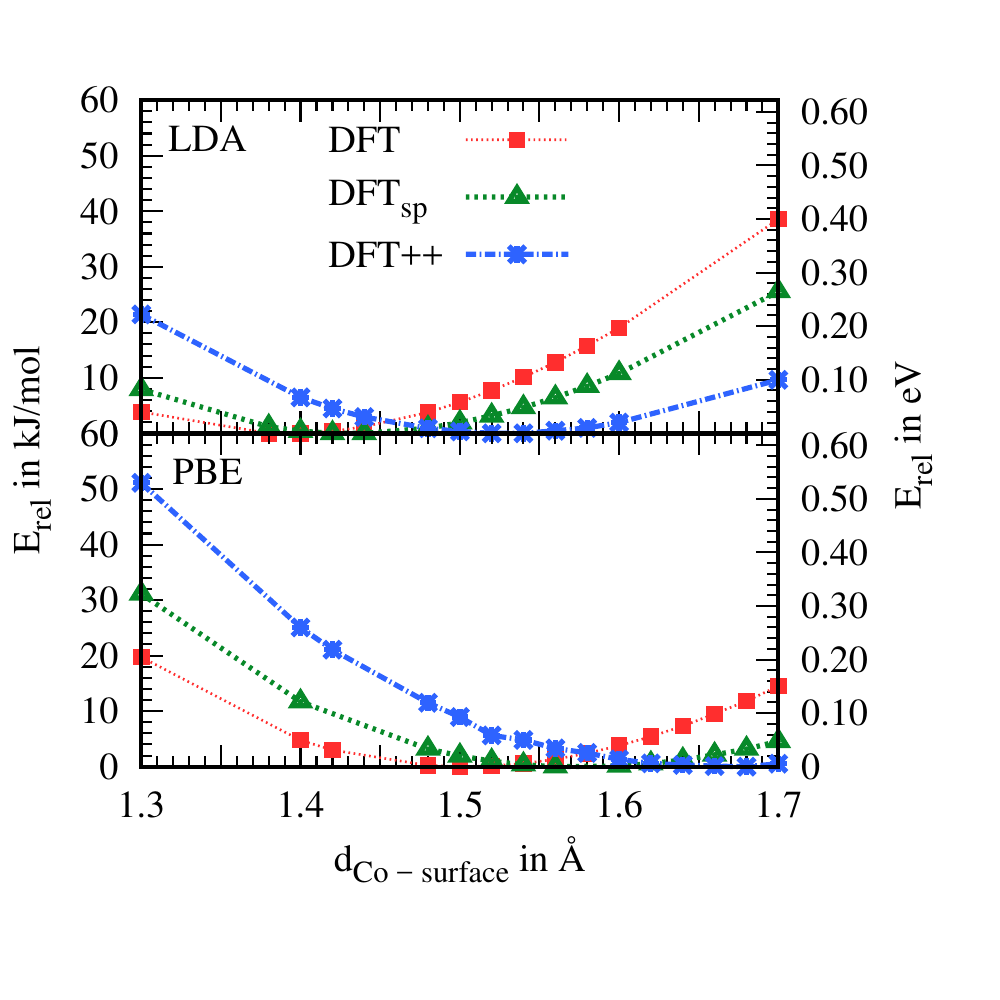}
\caption{Relative energies obtained from spin-unpolarized DFT (DFT), spin-polarized DFT (DFT$_{\mathrm{SP}}$) and DFT++  for different adsorption distances of Co on Cu(001). The DFT++ calculations were performed at $\beta = 100$~eV$^{-1}$.  The DFT++ results for $U$ = 3.8 eV, 4.0 eV and 4.2 eV are nearly on top of each other (see Supplemental Material), which is why we limit this plot to the results of $U$ = 4.0 eV. The exchange parameter for all values of $U$ is $J$ = 0.9~eV. For the definition of $d_{\mathrm{co-surface}}$ see Fig.\ \ref{unitcell}.}
\label{E_vs_d_100}
\end{figure}
\textcolor{changes}{The PES of a cobalt atom on a Cu(001) surface is studied with DFT (spin-polarized and spin-unpolarized) and DFT++, with the presently most used approximate exchange correlation functionals in solid state physics, namely LDA (LDA++) and PBE (PBE++). Furthermore, we provide the adsorption distance as obtained from DFT$+U$.} Although DFT++ is almost completely parametrized by the underlying DFT calculation, the Coulomb matrix (last term in  Eq.\ (\ref{siam_hamilton})) and the double-counting energy $E_{\mathrm{DC}}$ are in practice unknown, although efforts to compute the Coulomb matrix \cite{jaco15} as well as the double-counting \cite{Haule2015} from first-principles exist.
The variation of the latter can be used to control the occupation~\footnote{Unless the exact exchange correlation functional is known, the exact occupation is unknown too.} on the correlated atom (the Co atom in our case) manually (as, e.g. in Ref. \onlinecite{karolak14}), but here we need a unique way to define the double-counting at each adsorption distance, which is achieved by applying the FLL double-counting correction (Eq.\ (\ref{fll_dc})). Of course, the same argumentation holds for choosing $U$. As mentioned before, in this study we address this issue by comparing three different fixed values for $U$ ($U = 3.8,~4.0$ and $4.2$ eV) and one fixed value for $J$ (0.9 eV) at each data point, allowing for the investigation of the PES with increasing $U$ (for the parametrization details see Sec.\ \ref{method}).
\textcolor{changes}{In addition, we checked for $U$ = 3.0~eV and $U$ = 5.0~eV if the adsorption distance (LDA++ only) changes for these values, by computing only a few points around the LDA++ ($U$ = 4.0 eV) minimum of 1.52~\AA~(see Supplemental Material)}, \textcolor{changes4}{suggesting that slightly varying $U$ around 4.0~eV has no large impact on the adsorption distance.}\\
The total energies as a function of the adsorption distance for DFT and DFT++ ($\beta = 100$~eV$^{-1} \sim 116$~K) are shown in Fig.\ \ref{E_vs_d_100} \textcolor{changes}{for the two functionals discussed here}. All energies were plotted relative to the lowest obtained energy for a given method (all minimum-energy adsorption distances are provided in Tab. \ref{d_min_table}).\\
\begin{table}[H]
 \caption{Adsorption distances in \AA~at which the energy minima were obtained within the different methods employed here. The values for $U$ are provided in the table and the value for $J$ is 0.9~eV.} 
\centering
 \begin{tabular}{c  c c }
 \hline\hline
& LDA & PBE  \\
\hline
DFT~ & 1.40  &  1.50 \\
DFT$_{\mathrm{sp}}$~ & 1.44  & 1.56 \\
DFT++ (3.0 eV)~ &1.52 & - \\
DFT++ (3.8 eV)~ &1.52 & 1.68 \\
DFT++ (4.0 eV)~ &1.52 & 1.68 \\ 
DFT++ (4.2 eV)~ & 1.54 &  1.68 \\
DFT++ (5.0 eV)~ & 1.54 &  - \\
DFT$+U$ (4.0 eV)~ & 1.53 & 1.69\\
\hline\hline
\end{tabular}
\label{d_min_table}
\end{table}  
As one can see in Tab. \ref{d_min_table}, the minima are shifted to higher adsorption distances in the series DFT $<$ DFT$_{\mathrm{sp}}$ $<$ DFT++/DFT$+U$. Remember that the ``DFT'' results are without spin polarization, thus the effect of including it (DFT$_{\mathrm{sp}}$) is a small shift towards higher adsorption distances by about 0.04~\AA~to 0.06~\AA. 
The effect of adding dynamical electron correlation (DFT++) continues shifting the PES towards larger adsorption distances: the shift from LDA$_{\mathrm{sp}}$ (LSDA) to LDA++ is about 0.08~\AA~to 0.10~\AA, and 0.12~\AA~ for PBE$_{\mathrm{SP}}$ to PBE++. For DFT$+U$ we have not calculated the PES for the surface adsorption, but rather optimized Co on Cu(001) with all Cu atoms kept frozen, which is why we provide the minima in Tab. \ref{d_min_table}. \textcolor{changes4}{The  shift of the adsorption minimum from LDA$_{\mathrm{SP}}$ to LDA+$U$ (1.44~\AA~$\rightarrow$ 1.53~\AA) is in agreement with LDA++, as it is for PBE$_{\mathrm{sp}}$ to PBE+$U$ (1.56~\AA~$\rightarrow$ 1.69~\AA) compared to PBE++.}\\
Generally, LDA (LDA$_{\mathrm{sp}}$) predicts a smaller adsorption distance than PBE (PBE$_{\mathrm{sp}}$) which is  not unexpected, because LDA  is known for its overbinding character \cite{Painter1982, Grossman1995, Mayor-Lopez1997}.  Accordingly, it is not surprising that we found the LDA$_{\mathrm{sp}}$ minimum 0.12~\AA~below the minimum of PBE$_{\mathrm{sp}}$. To account for dynamical electron correlation  within the DFT++ framework, however, does not solve this issue. Instead, it shifts the PESs of LDA and PBE towards higher adsorption distances, and further increases the difference between the minima for the two functionals (see Tab. \ref{d_min_table}). \textcolor{changes}{Further, the much cheaper DFT$+U$ method provides a similar answer as DFT++ concerning the minimum adsorption distance of Co/Cu(001).}
\textcolor{changes}{For understanding the shift introduced by DFT++ in contrast to DFT, we proceed by \textcolor{changes4}{separating} the correlation energy ($E_{\mathrm{pot}} - E_{\mathrm{DC}}$) introduced by CT-QMC from the total DFT++ energy ($E_{\mathrm{DFT++}}$, see Eq.\ (\ref{Edmft})). \textcolor{changes4}{The remaining terms should account for the isolated effect of the electron density being modified by adding explicit correlation.} $E_{\mathrm{pot}} - E_{\mathrm{DC}}$ as a function of the adsorption distance is shown in Fig.\ \ref{eint}. \textcolor{changes4}{It is negative and its absolute value increases with adsorption distance}.\
\begin{figure}[h]
\centering
\includegraphics[width=0.5\textwidth] {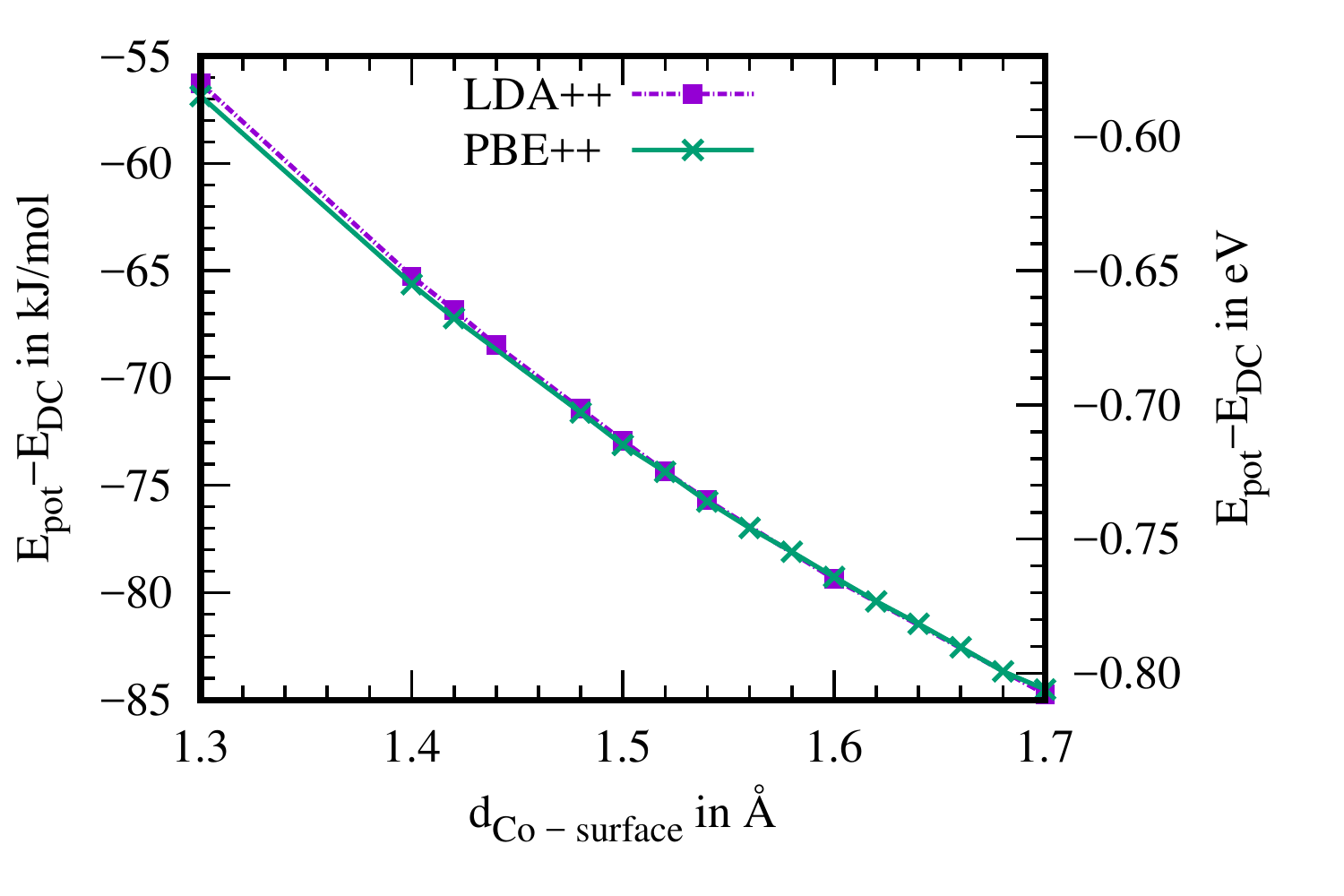}
\caption{$E_{\mathrm{pot}} - E_{\mathrm{DC}}$ as obtained from LDA++ and PBE++ at $\beta$ = 100 eV$^{-1}$ with $U$ = 4.0~eV and $J$ = 0.9~eV.}
\label{eint}
\end{figure}
The physical interpretation of the increasing $|E_{\mathrm{pot}} - E_{\mathrm{DC}}|$ with increasing distance is that the electrons on the Co $3d$ shell get more correlated at larger adsorption distances, \textcolor{changes5}{similar to H$_2$ being more statically correlated (in the chemistry sense) when stretched }. In contrast to the interaction term $E_{\mathrm{pot}} - E_{\mathrm{DC}}$, the remaining terms of Eq.\ (\ref{Edmft}) (shown in yellow in Fig.\ \ref{E_vs_d_100_decomp}) have a minimum close to the corresponding DFT$_{\mathrm{SP}}$ minimum, and from that the energy only slightly increases \textcolor{changes7}{with increasing adsorption distance} so that the decreasing $E_{\mathrm{pot}} - E_{\mathrm{DC}}$ term ``dominates'' and shifts the minimum towards a larger distance.}
\begin{figure} [b]
\centering
\includegraphics[width=0.5\textwidth]{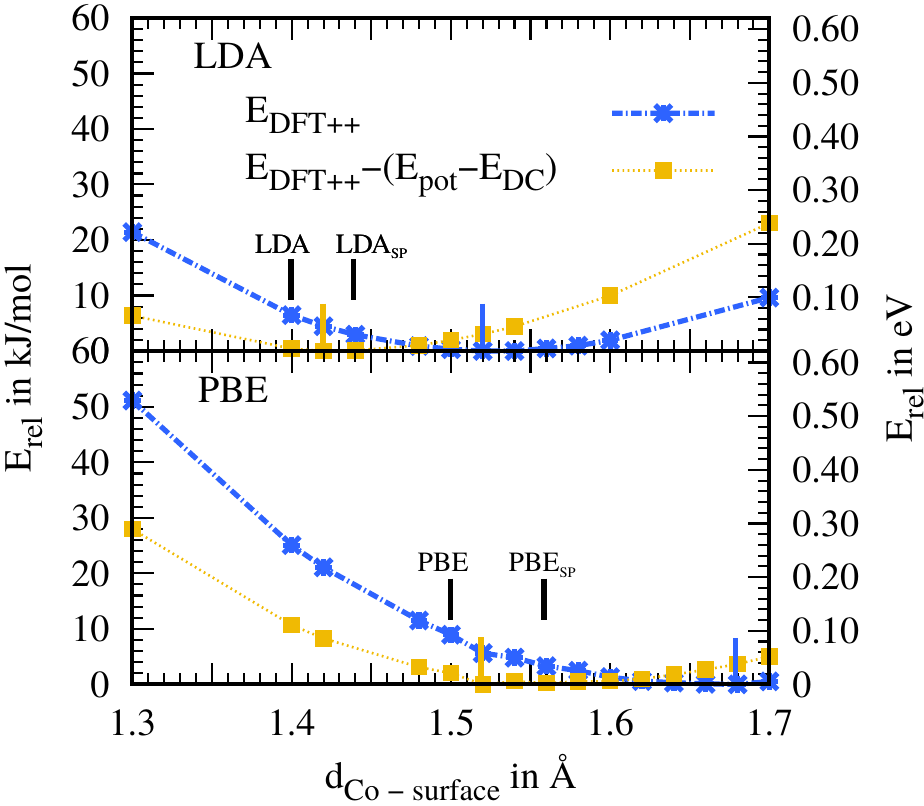}
\caption{Relative energy obtained from  DFT++  for different adsorption distances of Co on Cu(001) and for $E_\mathrm{DFT++}-(E_{\mathrm{pot}} - E_{\mathrm{DC}})$. Results obtained  at $\beta = 100$~eV$^{-1}$ with $U$ = 4.0~eV and $J$ = 0.9~eV. Bars in black marking the adsorption distances for DFT/DFT$_{\mathrm{SP}}$, and colored bars marking the DFT++ ($U$ = 4.0 eV) respectively the $E_{\mathrm{DFT++}}$ - $(E_{\mathrm{pot}} - E_{\mathrm{DC}})$ adsorption distance.}  
\label{E_vs_d_100_decomp}
\end{figure}
\subsection{Electronic structure of Co/Cu(001)}\label{qmc}
In the previous section, we could show that there is a non-negligible shift of the PES minimum by taking into account \textcolor{changes2}{the dynamical effects within the DFT++ framework}, \textcolor{changes}{as well as \textcolor{changes2}{by adding a Coulomb potential in the framework of DFT$+U$}}. Here we would like to gain insight into how different adsorption distances affect the electronic structure of an adsorbate.\\
\begin{figure}
\centering
\includegraphics[width=0.5\textwidth] {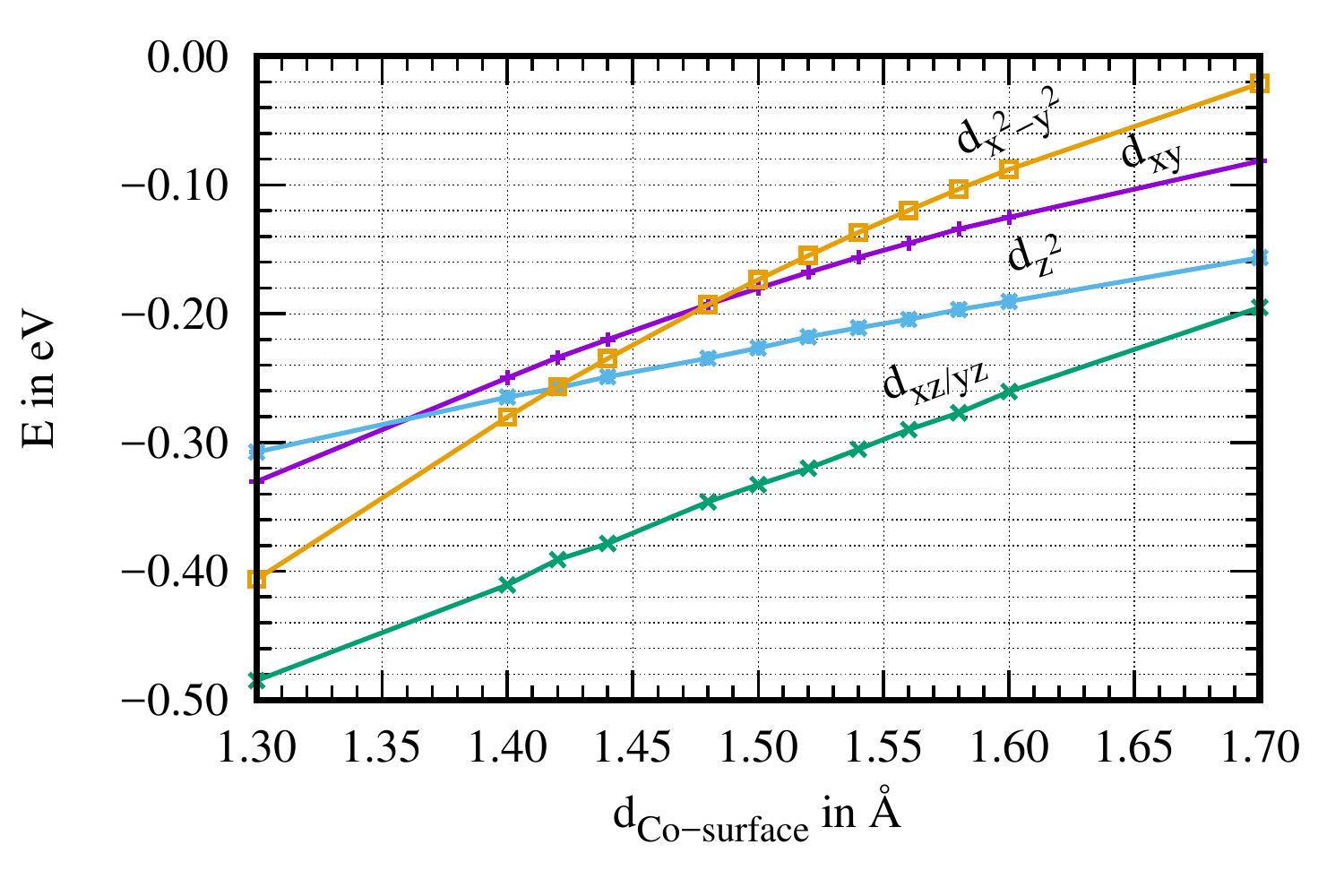}
\caption{Static crystal-field splitting (i.e. $3d$-orbital energies of Co) of the Co $3d$ shell as obtained from LDA (spin-unpolarized), as a function of the adsorption distance. \textcolor{changes4}{The Fermi level is set to zero}.}
\label{dftlevels}
\end{figure}
Besides the parametrization of the Coulomb matrix (discussed in Sec.\ \ref{method}), it is important for this discussion to have a look at the hybridization functions and impurity levels (the crystal-field splitting) at different adsorption distances, because these parameters \textcolor{changes7}{as obtained from DFT} are those entering the AIM. However, to keep this discussion concise, we will focus on the most relevant orbitals, namely s which are singly occupied, \textcolor{changes}{and thus the most promising for \textcolor{changes4}{rationalizing} the experimentally observed Kondo effect}. By inspection of Tab. \ref{lda_occupation}, one can identify the $d_{xy}$ and the $d_{z^2}$ orbital to be the orbitals of interest \textcolor{changes7}{ which is in agreement with Refs. \onlinecite{Huang2008}  and \onlinecite{jaco15}\footnote{Note the different orientation of the $x$- and $y$- axis.}}. The energy level of each Co $3d$ orbital \textcolor{changes4}{as observed from spin-unpolarized LDA with respect to the Fermi level (0 eV) is shown in Fig.\ \ref{dftlevels}},  \textcolor{changes5}{where the energetic difference between the Co $3d$ levels can be seen as the crystal field splitting. \textcolor{changes7}{The results of PBE are qualitatively and \textcolor{changes6}{almost} quantitatively in agreement with LDA (see Supplemental Material)}}. One can notice that with increasing adsorption distance the crystal-field splitting changes qualitatively and quantitatively, e.g., for adsorption distances between 1.35~\AA~ and 1.50~\AA, the energetic ordering of $d_{xy}$, $d_{x^2-y^2}$ and $d_{z^2}$ changes \textcolor{changes3}{multiple times}.\\ 
\begin{table*}
\centering
 \caption{Occupation of the Co $3d$ shell as obtained from LDA++ at $\beta = 100$ eV$^{-1}$ and $U$ = 4.0~eV ($J$ = 0.9~eV) via integration of the interacting impurity Green's function $G_{\mathrm{IMP}}$, at selected values for the adsorption distance. All values were obtained by setting the double-counting as calculated from the fully localized limit. Values in parentheses are obtained from integration of the non-interacting Green's function (Eq.\ (\ref{nonint_imp})). \textcolor{changes6}{ We also show the expectation value of the spin, as well as the weight $w$ of the most probable local many-body configuration, which is $d^8$, $S$=1 with half-filled $d_{xy}$ and $d_{z^2}$. See Supplemental Material for the full state histogram.}} 
\begin{tabular}{c c  c  c  c  c c c}
\hline\hline

$d_{\mathrm{Co-surf.}}$ & $d_{xy}$ & $d_{xz/yz}$ &$d_{z^2}$ & $d_{x^2-y^2}$ & total & $\langle S_z\rangle$ & $w$ \\
\hline
1.3~\AA & 1.08 (1.45)& 1.86 (1.57)& 1.07 (1.56)& 1.86  (1.72) &  7.73 (7.87) & 0.95 & 0.54\\
1.5~\AA & 1.04 (1.47)& 1.88 (1.56)& 1.05 (1.57)& 1.88 (1.71)& 7.73    (7.87) & 0.99 & 0.61\\
1.7~\AA & 1.02 (1.47)& 1.89 (1.55)& 1.03 (1.58)&  1.90 (1.70)& 7.73   (7.85) & 1.01 & 0.65\\
\hline\hline
\end{tabular}

\label{lda_occupation}
\end{table*}  
Regardless of the adsorption distance, the $d_{xy}$ and the $d_{z^2}$ orbitals are singly occupied (Tab. \ref{lda_occupation}), although above 1.48~\AA~the  $d_{x^2-y^2}$ orbital is highest in energy. We emphasize here that this observation is not a violation of the Aufbau principle, which is fulfilled for the Kohn--Sham Bloch wavefunction. Instead, the Co $3d$ orbitals whose occupations are reported here result from projection onto local orbitals, as mentioned in Sec.\ \ref{method}. \textcolor{changes6}{This \textcolor{changes7}{and the parametrization of the Coulomb matrix} yields the atomic part of the Anderson Hamiltonian (last two terms of Eq.\ (\ref{siam_hamilton})), which in the hybridization expansion of CT-QMC is solved first. We can follow the change of the atomic ground state from $d^8$, $S$=1 with half-filled $d_{xy}$ and $d_{z^2}$ at smaller distances to $d^8$, $S$=1 with half-filled $d_{x^2-y^2}$ and $d_{z^2}$ at larger distances (as intuitively expected by inspection of Fig.\ \ref{dftlevels}). In the course of the 
QMC calculation the dynamical hybridization $\Delta(\omega)$ (for its real part, see Fig. 4 of the Supplemental Material) is evaluated in a formal perturbation theory to all orders, and precisely this dynamics ensures that for the complete Hamiltonian the local configuration with the largest contribution to the ground state is $d^8$, $S$=1 with half-filled $d_{xy}$ and $d_{z^2}$ at all distances. In Tab.\ \ref{lda_occupation} we show the contribution of said atomic state to the many-body ground state as well as the expectation value of $\hat{S}_z$. Both are consistent with the system becoming more decoupled from the surface and thus more atomic as the distance is increased (see Supplemental Material, Fig.\ 7 for a more detailed evaluation of the different charge and spin contributions)}.\\
\textcolor{changes2}{In the following, we consider the imaginary part of the hybridization function as obtained from LDA ($\mathrm{Im}\Delta$ obtained from PBE is provided in Fig.\ 3 of the Supplemental Material, and is in agreement with LDA}), which can be regarded as the dynamical broadening of the impurity levels, due to the hybridization with the Cu(001) surface. We focus on the half-filled orbitals (Fig.\ \ref{hybri}) and find that they retain their qualitative features when changing the adsorption distance, but as one would expect, the hybridization increases systematically as the adsorption distance is lowered. Fig.\ \ref{hybri} shows that this effect is a little bit more pronounced for the $d_{xy}$ orbital than for the $d_{z^2}$ orbital. Thus, we would expect that the Kondo effect in the $d_{xy}$ orbital is somewhat more sensitive towards changes in the adsorption distance than it is in the $d_{z^2}$ orbital. \textcolor{changes}{We can check this behavior} of $T_\mathrm{K}$ with the simplest Kondo-
model (one-band with a constant hybridization $\Gamma$) \cite{hewson} where the hybridization is related to the Kondo temperature as
\begin{equation}\label{simpleAIM}
k_bT_\mathrm{K} = \frac{\sqrt{\Gamma U}}{2}\mathrm{exp}\left(\frac{\pi\epsilon(\epsilon+U)}{\Gamma U}\right),
\end{equation}
 with $\epsilon$ being the energy level of the impurity \textcolor{changes4}{relative to the Fermi level}, $U$ the Coulomb interaction, $k_{\mathrm{b}}$ the Boltzmann constant and $\Gamma \sim -\mathrm{Im}\Delta(0)$.\\
  In Tab. \ref{TKs}, we have summarized the estimated Kondo temperatures for both the $d_{xy}$ and the $d_{z^2}$ orbital at different adsorption distances, \textcolor{changes4}{obtained with the Co $3d$-orbital energies ($\epsilon$) and hybridization strength ($\mathrm{Im}\Delta(0)$) from spin-unpolarized LDA}. Obviously, this estimation predicts too large Kondo temperatures for the $d_{xy}$ orbital. Although the hybridization at the Fermi level ($\omega$ = 0~eV) is strongly decreasing for larger adsorption distances, the Kondo temperature is increasing to unrealistic values (about 419.3~K at 1.70~\AA). 
The increasing Kondo temperature with increasing distance is a result of the impurity level being shifted towards the Fermi energy, thus \textcolor{changes3}{overcompensating} the decreasing hybridization. For the $d_{z^2}$ orbital, the Kondo temperature behaves qualitatively similar as for the $d_{xy}$ orbital. However, the obtained values for $T_K$ are in much better agreement with the experiment, and, as assumed earlier, are less dependent on the adsorption distance compared to the $d_{xy}$ orbital. \textcolor{changes4}{An estimation of the Kondo temperatures from spin-unpolarized PBE yields  a qualitatively similar picture as from LDA (see Supplemental Material)}.\\ \textcolor{changes6}{From our simple estimates of the Kondo temperatures, one observes the serendipitous nature of such simplistic assumptions for multi-orbital Kondo systems. In the present case, we would conclude that the $d_{xy}$ orbital would be dominant over $d_{z^2}$ in an underscreened or two-stage Kondo situation. However, a low temperature investigation\cite{jaco15}  using a perturbative solver suggests that it is in fact the other way around, the $d_{xy}$ Kondo temperature being strongly suppressed by charge fluctuations. We would like to mention here that these investigation  differs from ours in the way that it includes the effect of a tip, as well as a different description of the Cu(001) surface. \textcolor{PB}{In Ref. \onlinecite{jaco15}, the latter is modeled by a cluster approach embedded in a tight-binding Bethe lattice, whereas we applied periodic boundary conditions.}
}
  \begin{table}[H]
\centering
 \caption{Estimated Kondo temperature as obtained from a one-band model with a constant hybridization. \textcolor{changes}{$\mathrm{Im}\Delta(0)$ is the value of the imaginary part of the hybridization function at $\omega$ = 0~eV (Fermi level). $\epsilon_{xy/z^2}$ are the energies of the Co $d_{xy/z^2}$ orbitals relative to the Fermi level. Values taken here are obtained from LDA, and for $U$ we have chosen 4.0~eV. For the estimation of the Kondo temperature $T_{\mathrm{K}}$ see Eq.\ (\ref{siam_hamilton}). Experimental value for $T_{\mathrm{K}}$ = 88 $\pm 4$~K \cite{wahl02,wahl04,wahl09}. } } 
\begin{tabular}{c c  c  c  c  c c }
\hline\hline

$d_{\mathrm{Co-surf.}}$ & -Im$\Delta_{xy}$(0) & -Im$\Delta_{z^2}$(0) & $\epsilon_{xy}$ & $\epsilon_{z^2}$  &  $T_{\mathrm{K},xy}$ &  $T_{\mathrm{K},z^2}$\\
\midrule[0.75pt]
1.30~\AA & 0.258& 0.202& -0.330 & -0.308 &  147.7&   62.68 \\
1.40~\AA & 0.212& 0.181& -0.250 & -0.265&  165.7&  67.33   \\ 
1.50~\AA & 0.172& 0.160& -0.180& -0.227&  208.4&    69.32  \\
1.60~\AA & 0.139 & 0.140& -0.125 &  -0.190& 280.2&   75.82 \\
1.70~\AA & 0.112 & 0.122 & -0.081 &  -0.156& 419.3&  85.36 \\
\hline\hline
\end{tabular}
\label{TKs}
\end{table}  
\begin{figure}[h]
\centering
\includegraphics[width=0.48\textwidth] {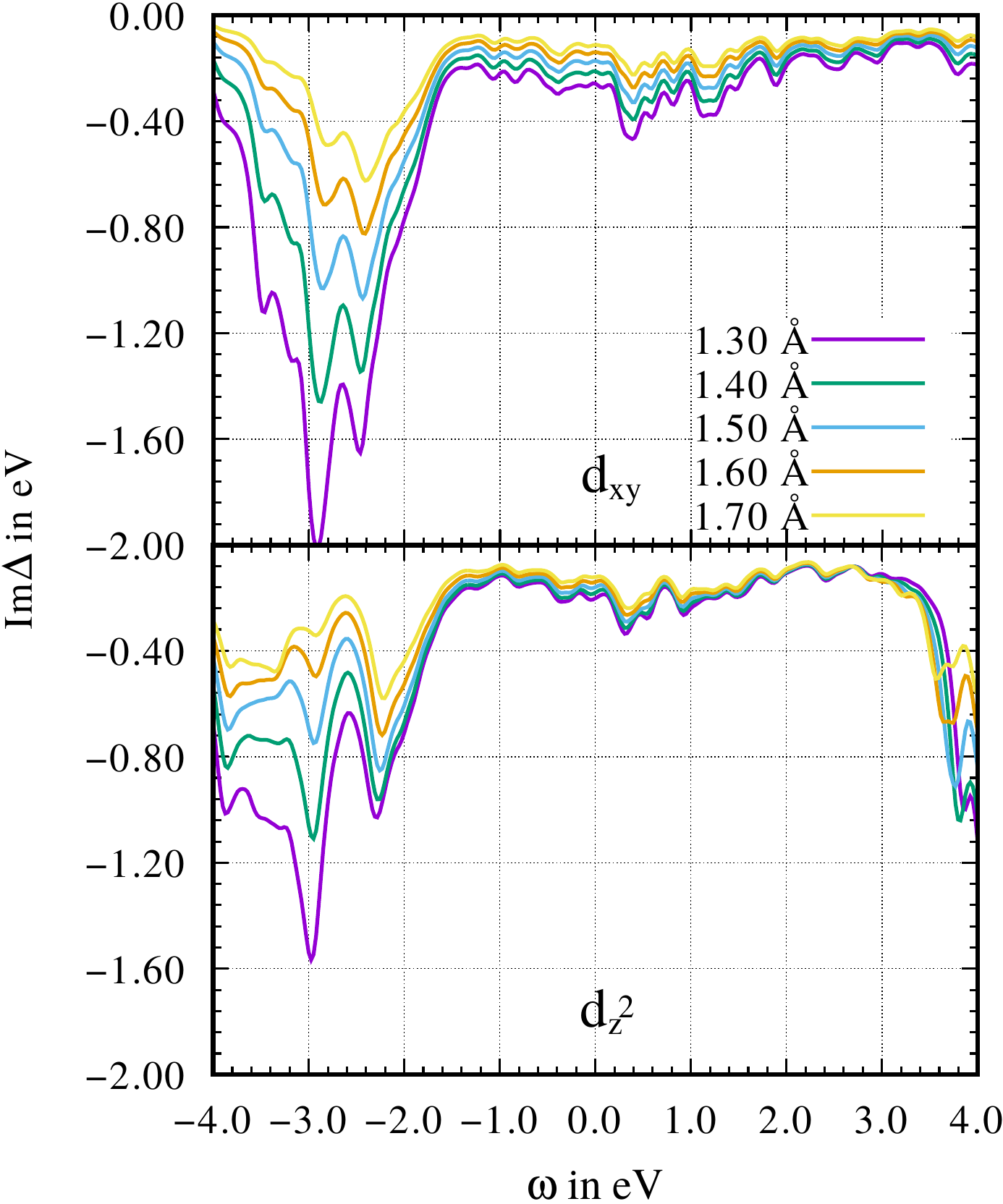}
\caption{Imaginary part of the hybridization function for the Co $3d_{xy}$ and the Co $3d_{z^2}$ orbital as obtained from LDA, for selected values of $d_{\mathrm{Co-surface}}$. Results for PBE are given in the Supplemental Material.}
\label{hybri}
\end{figure}
To complete our comparative study, we provide the self-energies  on the Matsubara axis \footnote{$\omega_n = \frac{(2n+1)\pi}{\beta}$, with $n$~= 0,1,2,...}  as obtained from DFT++ in Figure
 \ref{self}. This quantity is necessary for the interpretation of the electron correlation effects within the AIM. We limit the discussion to the half-filled orbitals only. From Fig.\ \ref{self} a) and b), one can see by comparing the LDA++ with the PBE++ results, that both electronic structure methods agree well with each other at all selected adsorption distances.  This can be understood as follow: the parametrization for the AIM is completely done by DFT, as introduced in Sec.\ \ref{method}, which yields the energy-dependent hybridization function (Eq.\ (\ref{hyb_function})), describing the coupling of the Co $3d$ shell to a non-interacting ``bath'' of electrons (the Cu(001) surface).
 Consequently, the agreement between both functionals is due to the similar description of the coupling of the Co $3d$ shell to the Cu(001) surface at a given adsorption distance (\textcolor{changes3}{compare Fig.\ \ref{hybri} and  Fig.\ 3 of the Supplemental Material}). By further inspection of Fig.\ \ref{self}, one can see that we are not in the Fermi liquid regime (for $\beta = 100~\mathrm{eV}^{-1}$), because $\mathrm{Im}\Sigma(\omega_n)$ does not behave linearly as $\omega_n \rightarrow 0$. It is nonetheless worthwhile to consider Im$\Sigma$ at as a function of the adsorption distance, because the results of the AIM are potentially affected by the distance, due to changes in the hybridization function. Indeed, for both orbitals $\mathrm{Im}\Sigma$ drops to larger absolute values as the distance is increased, which means that the electrons are ``more'' correlated at larger adsorption distances, \textcolor{changes5}{and is thus in agreement with the increasing term $|E_{\mathrm{pot}} - E_{\mathrm{DC}}|$ (
Fig.\ \ref{eint}) discussed before. In contrast, the electron correlation in the almost fully occupied orbitals (Co $3d_{xz/yz}$ and $3d_{x^2-y^2}$) are only little affected by the adsorption distances, as can be seen from $\mathrm{Im}\Sigma$ shown in Fig.\ 5 of the Supplemental Material (note the reduced abscissa). Consequently, the double occupancies showing the strongest variation as a function of the distance are  $\langle \hat{n}_{xy\sigma} \hat{n}_{xy\sigma'}\rangle$ , $\langle \hat{n}_{z^2\sigma}\hat{n}_{z^2\sigma'}\rangle$ and $\langle \hat{n}_{xy\sigma}\hat{n}_{z^2\sigma'}\rangle$. While all other terms change only by at most $\pm$5 \% from 1.3~\AA~to 1.68~\AA, these terms decrease by 57-68 \%. This leads to the conclusion that the shifts of the minima discussed in Sec.\ \ref{pes} mainly originates from both the Co $3d_{xy}$ and the Co $3d_{z^2}$ orbital.}
 \begin{figure}[h]
\centering
\includegraphics[width=0.48\textwidth] {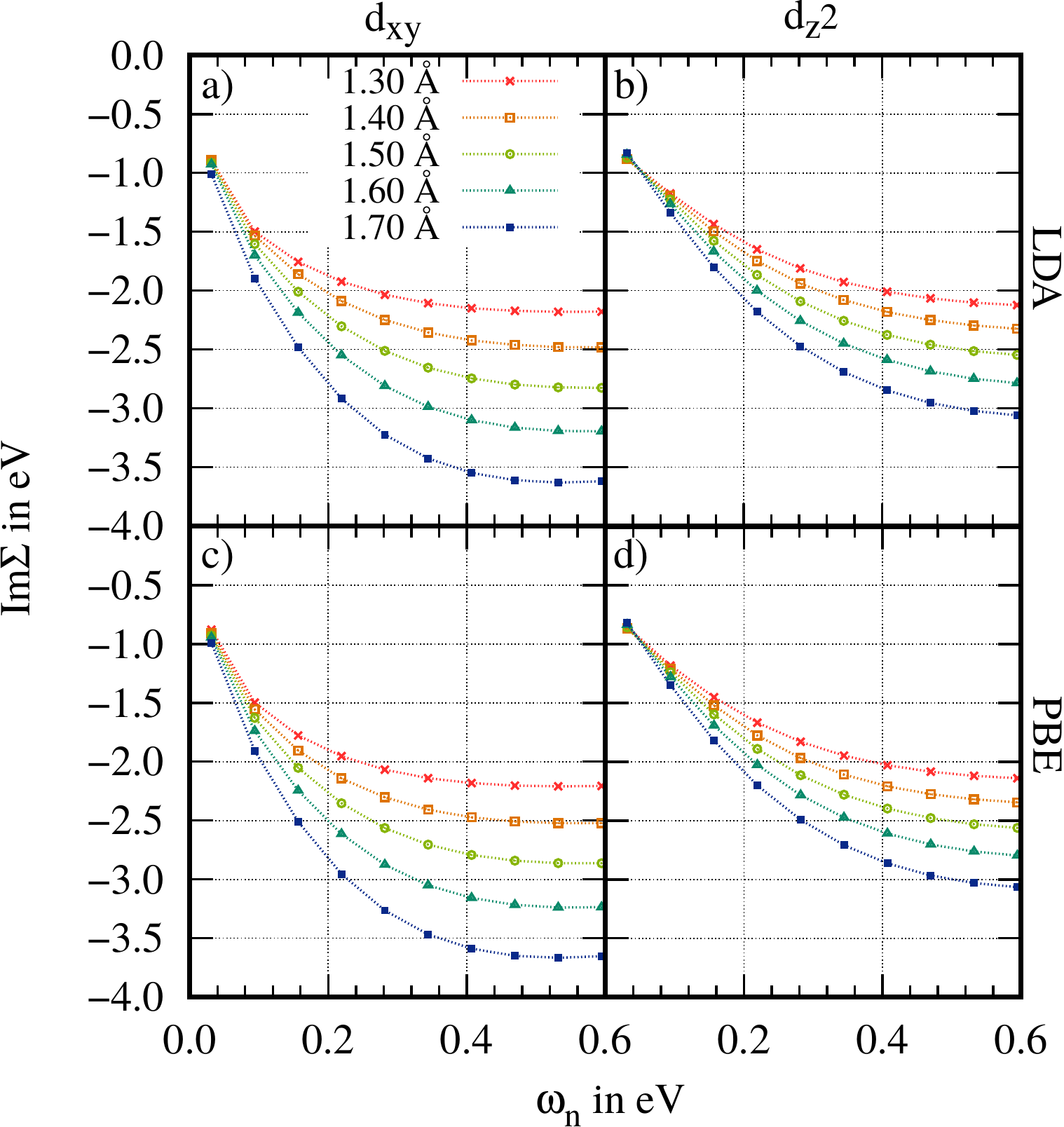}
\caption{Self energies (last term of Eq.\ (\ref{dyson})) obtained from DFT++ at $\beta = 100$~eV$^{-1}$ with $U = 4.0$~eV and $J = 0.9$~eV. a) Im$\Sigma$ for the Co $3d_{xy}$ orbital as obtained from LDA++ and PBE++. b) Im$\Sigma$ for the Co $3d_{z^2}$ orbital as obtained from LDA++ and PBE++.}
\label{self}
\end{figure}
 \section{Conclusion}\label{concl}
%
In summary, we could show that for Co/Cu(001) adsorption distances increases when applying the AIM on top of spin-unpolarized DFT by about 0.14~\AA~- 0.23~\AA, and 0.08~\AA~-0.12~\AA~when comparing with spin-polarized DFT. \textcolor{changes4}{The shifts can be explained by a larger amount of correlation (thus reducing the total energy) at larger distances as suggested by our data. However, this effect is already captured by DFT$+U$, which is much cheaper than DFT++.} On the other hand, given that the differences in adsorption distance between DFT and DFT++/DFT$+U$ are of the same order as the DFT error bar 
\footnote{\textcolor{PB}{Although Kohn--Sham density functional theory is formally correct, the exact exchange--correlation functional is still unknown, for which reason one has to rely on present-day approximate exchange--correlation functionals. The approximate nature of these functionals causes an error compared to experimental values (e.g. for lattice parameters and adsorption distances), as comprehensively shown in Ref. \onlinecite{Heyd2004, Wu2006, Ma2011}}}, 
this implies that the effect of strong correlation on these distances could also be adequately described by a suitable chosen approximate exchange--correlation functional within DFT.\\
Further, for the self-energies \textcolor{PB}{$\Sigma$} as obtained from DFT++, it is less important whether the AIM is parametrized by LDA or PBE. However, \textcolor{changes5}{$\mathrm{Im}\Sigma$ delicately depends on the adsorption distance of Co on Cu(001) for the Co $d_{xy}$ and $d_{z^2}$ orbitals, implying that the increasing correlation effects (with increasing distance) in these orbitals are mainly responsible for the observed shifts.}

\textcolor{PB}{One might argue that a larger distance from the surface
could imply smaller energy differences between different adsorption sites,
and between adsorption sites and diffusion transition states,
possibly leading to faster diffusion.
This relative effect of strong correlation on these quantities is not
straightforward to predict and might be an interesting direction for future research.}
 \section{Acknowledgment}
The authors acknowledge the high-performance-computing team of the \textit{Regionales Rechenzentrum: Universit\"at Hamburg} and the \textit{North-German Supercomputing Alliance (HLRN)} for technical support and computational resources. M. K. acknowledges financial support
by the \textit{Deutsche Forschungsgemeinschaft} (DFG) through SFB 1170.\\
 \section{Appendix}\label{appendix}
 \subsection{Dynamic\textcolor{changes5}{al} vs. static correlation}\label{dyn_stat}

\textcolor{changes2}{We would like to explain the terms "dynamical" and "static/non-dynamical" as they are used in the fields of solid-state physics and quantum chemistry. In quantum chemistry, the term ``dynamical'' electron correlation refers to electrons avoiding each others due to their Coulomb repulsion. The ``non-dynamical'' or ``static'' correlation refers to a situation where the ground state of a system can not be described by a single Slater determinant, as in the case where different \textcolor{changes4}{configurations} (Slater determinants) are close in energy, or are degenerate.\\
In physics, the term ``dynamical'' is used for  frequency- or energy-dependent quantities such as, e.g., when the self-energy or the hybridization function are not constant in energy. ``Static'' in this sense means the opposite situation, e.g., in Hartree-Fock theory the self-energy is a pure energy/frequency-independent term. When referring to DFT$+U$ to as being non-dynamical/statical, we mean that the self-energy in this approach is still frequency independent (\textcolor{changes3}{although DFT does include what chemists call dynamical correlation, i.e., including the effect of electrons avoiding each other due to their Coulomb repulsion}), because the added Coulomb potential is a mean-field like term \cite{Lichtenstein1998}.}

\pagebreak

\end{document}